\documentclass[11pt]{article}
\usepackage{latexsym,amssymb,epsf}

\begin{document}

\def\bea{\begin{eqnarray}}
\def\eea{\end{eqnarray}}
\def\be{\begin{equation}}
\def\ee{\end{equation}}
\def\nn{\nonumber\\}
\renewcommand{\theequation}{\mbox{\arabic{section}.\arabic{equation}}}

\title{Quantized bulk scalar fields in the Randall-Sundrum brane model}

\author{A.~Flachi\thanks{e-mail address:{\tt antonino.flachi@ncl.ac.uk}}\ \ and D.~J.~Toms\thanks{e-mail address:{\tt d.j.toms@newcastle.ac.uk}}\\
Department of Physics, University of Newcastle upon Tyne,\\
Newcastle Upon Tyne, United Kingdom NE1 7RU}

\date{\today}
\maketitle
\begin{abstract}We examine the lowest order quantum corrections to the effective action arising from a quantized real scalar 
field in the Randall-Sundrum background spacetime. The leading term is the familiar vacuum, or Casimir, energy density. 
The next term represents an induced gravity term that can renormalize the 4-dimensional Newtonian gravitational constant. 
The calculations are performed for an arbitrary spacetime dimension. Two inequivalent boundary conditions, corresponding to 
twisted and untwisted field configurations, are considered. A careful discussion of the regularization and renormalization of 
the effective action is given, with the relevant counterterms found. It is shown that the requirement of self-consistency of 
the Randall-Sundrum solution is not simply a matter of minimizing the Casimir energy density. The massless, conformally coupled 
scalar field results are obtained as a special limiting case of our results. We clarify a number of differences with previous work.
\end{abstract}
PACS number(s):11.10.Kk, 04.50.+h, 04.62.+v, 11.25.Mj
\eject

\section{Introduction}
The search for a unified field theory led from the very early days after Einstein's theory of general relativity to the suggestion that there could be extra spatial dimensions beyond the familiar three. The original idea of Kaluza \cite{Kaluza} was to incorporate electromagnetism into a 5-dimensional theory of gravity. In order to explain why the extra spatial dimension has not been observed, Klein \cite{Klein} suggested that it was compactified to a circle of extremely small radius, typically taken to be of the order of the Planck length. The generalization to more than one extra spatial dimension \cite{DeWitt} allowed the incorporation of non-abelian gauge fields as part of the higher dimensional metric. These higher dimensional theories are usually referred to by the generic term Kaluza-Klein theories.

In the early 1980's an immense amount of interest was generated in Kaluza-Klein theories by Witten's \cite{Witten} observation that seven extra spatial dimensions was the minimum required for incorporation of an $SU(3)\times SU(2)\times U(1)$ gauge symmetry, and Nahm's proof \cite{Nahm} that this was the maximum allowed by supergravity. This focused attention on 11-dimensional theories with various compactifications. (See Ref.~\cite{DuffNilssonPope} for a review.) Further interest in higher dimensional theories was, and continues to be, generated by superstring theory.

A new angle on Kaluza-Klein theories was pointed out recently \cite{Arkani}. In contrast to the standard belief that extra dimensions must be associated with extremely small length scales, it was noted that the extra dimensions could be much bigger than the Planck scale and still remain as yet undetected. This has the effect of bringing the fundamental Planck scale much closer to the electroweak scale, and thus offers a possible solution to the gauge hierarchy problem. The basic idea behind this \cite{Arkani} is to have a $(4+n)$-dimensional spacetime, with the $n$-dimensional space having a volume $V_n$. The relationship between the 4-dimensional Planck scale $M_{Pl}$ and the fundamental mass scale $M$ of the higher dimensional space is $M_{Pl}\sim V_nM^{n+2}$. For large volumes $V_n$, we can have $M$ of the order of the electroweak scale $(\sim1$~TeV) and still have $M_{Pl}\sim10^{19}$~GeV. Unfortunately this scenario does not provide a completely satisfactory solution to the gauge hierarchy problem since it trades a large ratio between the Planck scale and the electroweak scale for a large ratio between the compactification scale $V^{-1/n}$ and the electroweak scale. 

Randall and Sundrum \cite {RS} suggested a model that overcomes the problem just described. This model is based on a 5-dimensional spacetime with the extra spatial dimension having an orbifold compactification. Two 3-branes with opposite tensions sit at the orbifold fixed points. The line element is
\be
ds^2=e^{-2kr|\phi|}\eta_{\mu\nu}dx^\mu dx^\nu-r^2d\phi^2\label{1.1}
\ee
with $x^\mu$ the usual 4-dimensional coordinates, $|\phi|\le\pi$ with the points $(x^\mu,\phi)$ and $(x^\mu,-\phi)$ identified. The factor of $e^{-2kr|\phi|}$ in the line element is often referred to as the warp factor, since it results in the spacetime not being the direct product of 4-dimensional spacetime and the extra dimension. The 3-branes sit at $\phi=0$ and $\phi=\pi$. $k$ is a constant of the order of the Planck scale (the natural scale for the theory), and $r$ is an arbitrary constant associated with the size of the extra dimension.

The interesting feature of the Randall-Sundrum model is that it can generate a TeV mass scale from the Planck scale in the higher dimensional theory. A field with a mass $m_0$ on the $\phi=\pi$ brane will have a physical mass of $m\simeq e^{-\pi kr}m_0$. By taking $kr\simeq12$, and $m_0\simeq10^{19}$~GeV, we end up with $m\simeq1$~TeV. In the original version of the Randall-Sundrum model all of the standard model particles are supposed to be confined on the brane, with only gravity in the bulk (5-dimensional) spacetime. Alternatives to confining particles to the brane, and allowing fields to live in the bulk spacetime have been investigated. See for example \cite{GW1,DHR,Pomarol,Grossman,Kitano,Chang,Gherghetta,Changetal,DHR2,Huber,Flachi}. It is the possibility of bulk matter fields that we are interested in this paper.

The radius $r$ is assumed to be the vacuum expectation value of a scalar field, called the radion. In the original Randall-Sundrum model the value of $r$ is undetermined. In order to make the theory physically acceptable a mechanism for determining $r$ must be found. In Ref.~\cite{GW2} a classical method for stabilizing $r$ was suggested.

The similar problem of fixing the size of the extra dimensions was found in the older Kaluza-Klein theories that were based on direct products of Minkowski spacetime and a homogeneous space, often chosen to be a sphere. It was necessary to have the extra dimensions of the order of the Planck scale, but there was nothing in the classical theory that determined this. It was recognized by Candelas and Weinberg \cite{CW} that quantum effects from matter fields or gravity could fix the size of the extra dimensions in a natural way. It therefore is sensible to ask if a similar procedure can fix the value of $r$ in the Randall-Sundrum model. This problem has received some recent attention \cite{Garriga,DJTPLB,GR} with some differing conclusions. The main purpose of the present paper is to provide details of what we believe to be the correct and complete calculation of the vacuum energy density for bulk scalar fields, and to calculate the induced gravity term in the effective action \cite{DJTgravity} since it is known \cite{CW} that this can play an important role in self-consistent solutions and can give different conclusions from merely considering the vacuum energy density. We also clarify the two inequivalent field configurations which can arise as a result of the boundary conditions that can be imposed. Finally, we analyze in detail the requirement that the Randall-Sundrum spacetime be a solution to the quantum corrected Einstein equations. It is shown that this is not just the condition that the Casimir energy density have a minimum.

\section{Quantum corrections to the classical theory}
\label{sec-quantum}
\setcounter{equation}{0}

In order to calculate the quantum corrections to the classical theory we will follow the procedure used in Ref.~\cite{DJTgravity}. This consists of first expanding the higher dimensional field in terms of a complete set of modes appropriate to the extra dimensions. The dependence on the extra dimensions can then be integrated out of the  action functional to leave an equivalent 4-dimensional theory, but one with an infinite number of fields. The masses of the fields are usually quantized in some way. This is the procedure used in our computation of the vacuum energy density \cite{DJTPLB} for the minimally coupled scalar field.

If we wish to calculate the induced gravity contribution to the effective action we must generalize the metric (\ref{1.1}) to allow curvature in the 4-dimensional submanifold representing the branes. We will take
\be
ds^2=e^{-2\sigma(y)}g_{\mu\nu}(x)dx^\mu dx^\nu-dy^2\label{2.1}
\ee
where $y$ is related to the angle $\phi$ in (\ref{1.1}) by
\be
y=r\phi\label{2.2}
\ee
and the function $\sigma(y)$ is given by
\be
\sigma(y)=k|y|\;.\label{2.3}
\ee
We will allow the submanifold specified by $y=0$ to have an arbitrary dimension $D$. (The Randall-Sundrum model is then specified by $D=4$ and $g_{\mu\nu}=\eta_{\mu\nu}$.)

The Einstein-Hilbert gravitational action is
\be
S_G=(16\pi G)^{-1}\int d^Dxdy\,|\hat{g}|^{1/2}\,(\hat{R}-2\Lambda)\label{2.4}
\ee
where we use a caret to denote a $(D+1)$-dimensional object. (So $\hat{g}_{\hat{\mu}\hat{\nu}}$ is the $(D+1)$-dimensional metric found from (\ref{2.1}), $\hat{g}$ is its determinant, and $\hat{R}$ is the scalar curvature constructed from it.) There are also terms \cite{RS} in the total action functional coming from the two branes at $y=0,\pi r$ which represent the hidden and visible branes respectively:
\be
S_{brane}=-\int d^Dxdy\,|\hat{g}|^{1/2}\,\left\lbrace V_v\delta(y-\pi r)+V_h\delta(y)\right\rbrace\;.\label{2.5}
\ee
We will return to this part of the action in Sec.~\ref{renormalization}.

In order that the metric in (\ref{2.1}) with $g_{\mu\nu}=\eta_{\mu\nu}$ satisfy the field equations coming from varying (\ref{2.4}) and (\ref{2.5}) with $\sigma(y)$ given in (\ref{2.3}) along with the $\mathbb{Z}_2$ identification $(x^\mu,y)\sim(x^\mu,-y)$, it follows that
\bea
\sigma''&=&2k\left\lbrack\delta(y)-\delta(y-\pi r)\right\rbrack\;,\label{2.6}\\
V_h=-V_v&=&\frac{(D-1)}{4\pi G}k\;,\label{2.7}\\
\Lambda&=&-\frac{1}{2}D(D-1)k^2\;.\label{2.8}
\eea
These reduce to the results of \cite{RS} when $D=4$.

For later use we record the curvature scalar for (\ref{2.1}):
\bea
\hat{R}&=&e^{2\sigma}R+D\left\lbrack2\sigma''-(D+1)(\sigma')^2\right\rbrack\label{2.9a}\\
&=&e^{2\sigma}R+4kD\left\lbrack\delta(y)-\delta(y-\pi r)\right\rbrack-D(D+1)k^2\;.\label{2.9b}
\eea

We will consider a single real scalar field $\Phi$ of mass $m$ in the spacetime (\ref{2.1}):
\be
S=\frac{1}{2}\int d^Dxdy\,|\hat{g}|^{1/2}\,\left\lbrace \hat{g}^{\hat{\mu}\hat{\nu}}\partial_{\hat{\mu}}\Phi\partial_{\hat{\nu}}\Phi-m^2\Phi^2-\xi\hat{R}\Phi^2\right\rbrace\;.\label{2.10}
\ee
We allow a general non-minimal coupling to the curvature with dimensionless coupling constant $\xi$. The minimally coupled scalar field has $\xi=0$. The conformally coupled scalar field has $\xi=(D-1)/(4D)$. We will keep $\xi$ arbitrary here.

By considering the $(D+1)$-dimensional field equation for $\Phi$, following the analysis of Ref.~\cite{GW1}, leads us to define $f_n(y)$ as a solution to
\bea
&&\hspace{-2cm}-e^{(D-2)\sigma}\partial_y\left\lbrack e^{-D\sigma}\partial_yf_n\right\rbrack+e^{-2\sigma}\bar{m}^2f_n\nn
&&+4\xi Dk e^{-2\sigma}\lbrack\delta(y)-\delta(y-\pi r)\rbrack f_n=m_n^2f_n\;,\label{2.11}
\eea
normalized by
\be
\int\limits_{-\pi r}^{+\pi r}dy\,e^{(2-D)\sigma}f_n(y)f_{n'}(y)=\delta_{nn'}\;.\label{2.12}
\ee
We have defined 
\be
\bar{m}^2=m^2-D(D+1)k^2\xi\;.\label{2.13}
\ee

We first obtain the boundary conditions satisfied by $f_n(y)$ by integrating (\ref{2.11}) first about $y=0$ and then about $y=\pi r$. We assume that $f_n(y)$ is continuous at these two points. This gives the two results
\bea
\lim_{\epsilon\rightarrow0}\left\lbrace \left.\frac{\partial f_n(y)}{\partial y}\right|_{y=\epsilon}- 
\left.\frac{\partial f_n(y)}{\partial y}\right|_{y=-\epsilon}\right\rbrace&=&4\xi kDf_n(0)\;,\label{2.14a}\\
\lim_{\epsilon\rightarrow0}\left\lbrace \left.\frac{\partial f_n(y)}{\partial y}\right|_{y=\pi r-\epsilon}- 
\left.\frac{\partial f_n(y)}{\partial y}\right|_{y=-\pi r-\epsilon}\right\rbrace&=&4\xi kDf_n(\pi r)\;,\label{2.14b}
\eea
(In the second result we have used the $y\sim -y$ identification.)

Next we solve (\ref{2.11}) for $y>0\ (y\ne\pi r)$ and $y<0\ (y\ne-\pi r)$. Call the solutions in these two regions $f^{\pm}_n(y)$. It is easy to show that
\be
f^{\pm}_n(y)=e^{\pm\frac{D}{2}ky}\left\lbrace A^{\pm}_nJ_\nu\left(\frac{m_n}{k}e^{\pm ky}\right) + B^{\pm}_nY_\nu\left(\frac{m_n}{k}e^{\pm ky}\right)\right\rbrace\label{2.14c}
\ee
where $A^{\pm}_n$ and $B^{\pm}_n$ are constants of integration. The order of the Bessel functions $\nu$ is given by
\be
\nu=\sqrt{\frac{\bar{m}^2}{k^2}+\frac{D^2}{4}}\;.\label{2.18}
\ee
At this stage we must choose how $f^{+}_n(y)$ is related to $f^{-}_n(y)$ by the ${\mathbb Z}_2$ identification on 
$S^1$. The situation is like that used some time ago \cite{Isham} in considering fields on non-simply connected 
spacetimes. For a real scalar field there are only two possibilities. We have either
\be
f_n(-y)=f_n(y)\;,\label{2.14d}
\ee
corresponding to an untwisted field, or else
\be
f_n(-y)=-f_n(y)\;,\label{2.14e}
\ee
corresponding to a twisted field. 
The first choice corresponds to the boundary 
conditions adopted in Ref.~\cite{GW1} and Ref.~\cite{Garriga}.  We examine each possibility.

For the untwisted field (\ref{2.14d}) leads to
\be
A^{-}_n=A^{+}_n\;,\ B^{-}_n=B^{+}_n\label{2.14f}
\ee
in which case we can write
\be
f_n(y)=e^{\frac{D}{2}k|y|}\left\lbrace A_nJ_\nu\left(\frac{m_n}{k}e^{ k|y|}\right) + B_nY_\nu\left(\frac{m_n}{k}e^{k|y|}\right)\right\rbrace \label{2.14g}
\ee
for either positive or negative values of $y$ (but $y\ne0,\pi r$), as given in Ref.~\cite{GW1}. Applying the boundary conditions (\ref{2.14a}) and (\ref{2.14b}) results in the eigenvalues $m_n$ determined by solutions to the transcendental equation
\be
F_\nu\left(\frac{m_n}{ka}\right)=0\;,\label{2.14}
\ee
with
\be
F_\nu(z)=\frac{2}{\pi}\left\lbrack y_\nu(az)j_\nu(z)-j_\nu(az)y_\nu(z)\right\rbrack\;,\label{2.15}
\ee
where $j_\nu(z)$ and $y_\nu(z)$ are expressed in terms of Bessel functions by
\bea
j_\nu(z)&=&\frac{1}{2}D(1-4\xi)J_\nu(z)+zJ_{\nu}^{\prime}(z)\;,\label{2.16}\\
y_\nu(z)&=&\frac{1}{2}D(1-4\xi)Y_\nu(z)+zY_{\nu}^{\prime}(z)\;.\label{2.17}
\eea
We have written
\be
a=e^{-k\pi r}\;.\label{2.19}
\ee
For $\xi=0$ and $D=4$ these expressions reduce to the results of Ref.~\cite{GW1}.

For the twisted field (\ref{2.14e}) leads to
\be
A^{-}_n=-A^{+}_n\;,\ B^{-}_n=-B^{+}_n\label{2.20a}
\ee
instead of (\ref{2.14f}). This time we have
\be
f_n(y)=\frac{y}{|y|}e^{\frac{D}{2}k|y|}\left\lbrace A_nJ_\nu\left(\frac{m_n}{k}e^{ k|y|}\right) + B_nY_\nu\left(\frac{m_n}{k}e^{k|y|}\right)\right\rbrace \label{2.20b1}
\ee
as the solution for $y>0\ (y\ne\pi r)$ and $y<0\ (y\ne-\pi r)$. Because $f_n(y)$ for the twisted field is an odd function of $y$, using the $y\sim-y$ identification, we must have
\be
f_n(0)=0\;,\ f_n(\pi r)=0\;.\label{2.20b}
\ee
These last two results are consistent with the general boundary conditions (\ref{2.14a}) and (\ref{2.14b}) since the derivative of an odd function is even, resulting in the left hand sides of the two equations vanishing. The boundary conditions (\ref{2.20b}) lead to a much simpler analysis, with the eigenvalues $m_n$ given as solutions to
\be
\tilde{F}_\nu\left(\frac{m_n}{ka}\right)=0\label{2.20c}
\ee
with
\be
\tilde{F}_\nu(z)=\frac{2}{\pi}\left\lbrack J_\nu(z)Y_\nu(az)-J_\nu(az)Y_\nu(z)\right\rbrack\;. \label{2.20d}
\ee
Thus there can be no conflict with the results of Ref.~\cite{DJTPLB} that 
used untwisted boundary conditions. 
As we discuss in Appendix B, there are some slight 
differences in the technicalities between the twisted and untwisted 
scalars.

We now expand the $(D+1)$-dimensional scalar field $\Phi(x,y)$ in terms of the $f_n(y)$ as
\be
\Phi(x,y)=\sum_n\varphi_n(x)f_n(y)\;.\label{2.20}
\ee
This expansion can be substituted into (\ref{2.10}) and the integration performed over $y$. The result is
\be
S=-\frac{1}{2}\sum_n\int d^Dx|g|^{1/2}\varphi_n(\Box+m_n^2+\xi R)\varphi_n\;.\label{2.21}
\ee
This shows that we may view the theory as one in $D$ dimensions, with an infinite number of fields whose masses are quantized according to (\ref{2.14}).

At this stage we can work out the one-loop effective action in the usual way: simply compute the contribution from the $n^{th}$ mode, and then sum over all modes. Because we are after the vacuum energy, as well as the induced gravity term, it is advantageous to use a heat kernel method, and we follow Ref.~\cite{DJTgravity} here. (See \cite{DJTChalkRiver} for a review.) 

The one-loop effective action for the $n^{th}$ mode is
\be
\Gamma^{(1)}_n=-\frac{i}{2}\int d^Dx|g|^{1/2}\int\limits_0^\infty\frac{ds}{s}\,e^{-ism_n^2}K(s;x,x)\label{2.22}
\ee
where
\bea
i\frac{\partial}{\partial s}K(s;x,x')&=&(\Box+\xi R)K(s;x,x')\;,\label{2.23}\\
K(0;x,x')&=&\delta(x,x')\;.\label{2.24}
\eea
The result for $\Gamma^{(1)}_n$ is infinite and requires regularization. We choose dimensional regularization here with $D$ viewed as $D+\epsilon$ with $\epsilon\rightarrow0$ understood. We now make use of the asymptotic expansion (see Ref.~\cite{DeWitt} for example)
\be
K(s;x,x)\simeq i\ell^\epsilon(4\pi is)^{-D/2}\sum_{k=0}^{\infty}(is)^ka_k\;,\label{2.25}
\ee
where
\bea
a_0&=&1\;,\label{2.26}\\
a_1&=&\left(\frac{1}{6}-\xi\right)R\;.\label{2.27}
\eea
Higher order terms in (\ref{2.25}) involve higher powers of the curvature tensor; so the $a_0$ and $a_1$ terms contain the leading order terms. 

The two leading terms in $\Gamma^{(1)}_n$ are found to be
\bea
\Gamma^{(1)}_n&\simeq&\frac{1}{2}(4\pi)^{-D/2}\ell^\epsilon\int d^Dx|g|^{1/2}\Big\lbrace\Gamma(-D/2)(m_n^2)^{D/2}\nn
&&+\Big(\frac{1}{6}-\xi\Big)\Gamma(1-D/2)(m_n^2)^{D/2-1}R\Big\rbrace\;.\label{2.28}
\eea
If we write
\bea
\Gamma^{(1)}&=&\sum_n\Gamma^{(1)}_n\nn
&=&\int d^Dx|g|^{1/2}\Big\lbrace {\mathcal L}^{(1)}_\Lambda+{\mathcal L}^{(1)}_R+\cdots\Big\rbrace\label{2.29}
\eea
we have
\bea
{\mathcal L}^{(1)}_\Lambda&=&\frac{1}{2}(4\pi)^{-D/2}\ell^\epsilon \Gamma(-D/2)\sum_n(m_n^2)^{D/2}\;, \label{2.30}\\
{\mathcal L}^{(1)}_R&=&\frac{1}{2}(4\pi)^{-D/2}\ell^\epsilon\Big(\frac{1}{6}-\xi\Big)R \Gamma(1-D/2)\sum_n(m_n^2)^{D/2-1}\;, \label{2.31}
\eea
The term in ${\mathcal L}^{(1)}_\Lambda$ represents a contribution to the cosmological constant in $D$ dimensions induced by quantum corrections to the classical theory. ${\mathcal L}^{(1)}_\Lambda$ is recognized as the negative of the Casimir, or vacuum energy density, and corresponds to what we calculated earlier \cite{DJTPLB} in the case $\xi=0$ and $D=4$. The term in ${\mathcal L}^{(1)}_R$ represents an induced gravity term which can renormalize the Newtonian gravitational constant found in the classical gravitational action. In the special case $\xi=1/6$, which corresponds to the conformal value for the coupling constant when $D=3$, we have ${\mathcal L}^{(1)}_R=0$. In this case the lowest order correction to the classical gravitational action would result in an $R^2$ theory. For the original 5-dimensional Randall-Sundrum metric, $D=4$, and the conformal value is $\xi=3/16$, so there is an induced gravity term in the effective action.

\section{Renormalization}
\label{renormalization}
\setcounter{equation}{0}

The expressions for ${\mathcal L}^{(1)}_\Lambda$ and ${\mathcal L}^{(1)}_R$ obtained in (\ref{2.30}) and (\ref{2.31}) may be evaluated using the method described in Appendix A or B. With $\omega(s)$ defined in (\ref{A1}) we have
\bea
{\mathcal L}^{(1)}_{\Lambda}&=&\frac{1}{2}(4\pi)^{-D/2}\ell^\epsilon(ka)^D\omega(-D/2)\;,\label{3.1}\\
{\mathcal L}^{(1)}_{R}&=&\frac{1}{2}(4\pi)^{-D/2}\ell^\epsilon \left(\frac{1}{6}-\xi\right)R(ka)^{D-2}\omega(1-D/2)\;.\label{3.2}
\eea
The pole parts of these expressions will determine the form of the required counterterms. If we use ${\rm PP}\lbrace\cdots\rbrace$ to denote the pole part of any expression, from (\ref{A43}) and (\ref{A44}) we find
\bea
{\rm PP}\left\lbrace{\mathcal L}^{(1)}_{\Lambda}\right\rbrace&=&\frac{1}{\epsilon}(4\pi)^{-D/2} \frac{k^D}{\Gamma(D/2)}\Big\lbrack a^D+(-1)^D\Big\rbrack d_D\;,\label{3.3}\\
{\rm PP}\left\lbrace{\mathcal L}^{(1)}_{R}\right\rbrace&=&\frac{1}{2\epsilon}(4\pi)^{-D/2} \left(\frac{1}{6}-\xi\right)R\;\frac{(D-2)k^{D-2}}{\Gamma(D/2)}\nn
&&\quad\times\Big\lbrack a^{D-2}+(-1)^D\Big\rbrack d_{D-2}\;.\label{3.4}
\eea
The first few coefficients $d_k$ are given in (\ref{A28a}--\ref{A28d}) for untwisted fields, and in (\ref{B5a}--\ref{B5f}) for twisted fields. In both cases higher order coefficients may be calculated in a routine manner, but because they become lengthy we will not give the results.

The renormalization can be effected by modifying the procedure used by Ref.~\cite{GR}, which used the brane tensions. Because we allow curvature on the branes, it is necessary to augment the brane tension terms with curvature or else it will not be possible to remove the pole terms coming from (\ref{3.4}) when $D$ is even. As noted in Ref.~\cite{GR}, the key feature is the $a$-dependence of the results. When $D$ is odd, (\ref{3.3}) and (\ref{3.4}) show that the pole terms are proportional to $a^D-1$ and $a^{D-2}-1$ respectively. Such pole terms can be dealt with by a renormalization of the bulk gravitational and cosmological terms. The sign difference between the factors of $a$ and the 1 comes about  from the integration over the extra $y$ coordinate to obtain the effective 4-dimensional theory. When $D$ is even ($D=4$ for the Randall-Sundrum model) the pole terms are proportional to $a^D+1$ and $a^{D-2}+1$ respectively. It is not possible to use the bulk terms to remove these divergences. The situation is easily understood from the viewpoint of heat-kernel methods and dimensional regularization, and was pointed out a long time ago \cite{DJTrenorm}. In an odd spacetime dimension (as occurs for $D=4$) the only possible counterterms come from the boundary of the spacetime. For the Randall-Sundrum model this corresponds to the two branes.  

The metric on the visible brane is
\be
g_{\mu\nu}^{v}(x)=\hat{g}_{\mu\nu}(x,y=\pi r)=a^2g_{\mu\nu}\;.\label{3.5}
\ee
The metric on the hidden brane is
\be
g_{\mu\nu}^{h}(x)=\hat{g}_{\mu\nu}(x,y=0)=g_{\mu\nu}\;.\label{3.6}
\ee
These give rise to 
\bea
R_v&=&a^{-2}R\;,\label{3.7}\\
R_h&=&R
\eea
as the curvature scalars on the visible and hidden branes We will take
\be
S_{brane}=S_{brane}^{v}+S_{brane}^{h}\label{3.9}
\ee
where
\bea
S_{brane}^{v}&=&\int\,d^Dx|g_v|^{1/2}(\kappa_vR_v-V_v)\;,\label{3.10}\\
S_{brane}^{h}&=&\int\,d^Dx|g_h|^{1/2}(\kappa_hR_h-V_h)\;,\label{3.11}
\eea
as the brane action. $\kappa_v,\kappa_h,V_v,V_h$ are all bare quantities. It follows that
\be
S_{brane}=\int\,d^Dx|g|^{1/2}\lbrace(\kappa_h+a^{D-2}\kappa_v)R-(V_h+a^DV_v)\rbrace\;.\label{3.12}
\ee

All bare quantities can be expressed in terms of renormalized ones and counterterms in the usual way~:
\bea
\kappa_{v,h}&=&\kappa_{v,h}^{R}+\delta\kappa_{v,h}\;,\label{3.13a}\\
V_{v,h}&=&V_{v,h}^{R}+\delta V_{v,h}\;,\label{3.13b}
\eea
This gives a counterterm brane action $\delta S_{brane}$ from (\ref{3.12}) in an obvious way. Because the structure of the pole terms in (\ref{3.3}) and (\ref{3.4}) is the same as that of $\delta S_{brane}$, there is no problem absorbing them into the counterterms.

There is always the freedom to perform finite renormalizations since the renormalization counterterms should be fixed by imposing some renormalization conditions on the effective action. (See Ref.~\cite{ColemanWeinberg} for example.) To this end we can write
\bea
\delta\kappa_{v,h}&=&\frac{\delta\kappa_{v,h}^{-1}}{\epsilon}+\delta\kappa_{v,h}^{0}\;,\label{3.14a}\\
\delta V_{v,h}&=&\frac{\delta V_{v,h}^{-1}}{\epsilon}+\delta V_{v,h}^{0}\;,\label{3.14b}
\eea
where $\delta\kappa_{v,h}^{-1},\delta\kappa_{v,h}^{0},\delta V_{v,h}^{-1},\delta V_{v,h}^{0}$ are all independent of $\epsilon$. Noting that the factors of $a^D$ in (\ref{3.12}) are really $a^{D+\epsilon}$, we find
\bea
\delta{\mathcal L}_{brane}&=&\Big\lbrack\frac{1}{\epsilon}\left(\delta\kappa_{h}^{-1}+a^{D-2}\delta\kappa_{v}^{-1}\right)\nn
&&\left(\delta\kappa_{h}^{0}+a^{D-2}\delta\kappa_{v}^{0}+a^{D-2}\delta\kappa_{v}^{-1}\ln a\right)\Big\rbrack\;R\nn
&&-\Big\lbrack\frac{1}{\epsilon}\left(\delta V_{h}^{-1}+a^D\delta V_{v}^{-1}\right)+\delta V_{h}^{0}\nn
&& + a^D\delta V_{v}^{0} + a^D\delta V_{v}^{-1}\ln a\Big\rbrack\label{3.15}
\eea
as the Lagrangian density counterterms on the branes. $D$ is now fixed to be the physical spacetime dimension ({\em eg}\/ 4). In order that the poles in (\ref{3.3}) and (\ref{3.4}) be removed we must fix
\bea
\delta V_{v}^{-1}&=&\left(\frac{k^2}{4\pi}\right)^{D/2}\frac{d_D}{\Gamma(D/2)}\;,\label{3.16a}\\
\delta V_{h}^{-1}&=&(-1)^D\delta V_{v}^{-1}\;,\label{3.16b}\\
\delta\kappa_{v}^{-1}&=&\frac{1}{2}\Big(\xi-\frac{1}{6}\Big)\frac{(D-2)k^{D-2}}{\Gamma(D/2)}\;d_{D-2}\;,\label{3.16c}\\
\delta\kappa_{h}^{-1}&=&(-1)^D\delta\kappa_{v}^{-1}\;.\label{3.16d}
\eea
If we use the results contained in the appendices for either twisted or untwisted fields, (see (\ref{A42},\ref{A43})), it can be seen that  ${\mathcal L}^{(1)}_{\Lambda}$ involves either terms independent of $a$, proportional to $a^D$ or $a^D\ln a$, apart from the term we have defined as $A_1$. Apart from $A_1$, the $a$-dependence is the same form as the finite counterterms $\delta V_{h}^{0}+a^D\delta V_{v}^{0}+a^D\delta V_{v}^{-1}\ln a$ appearing in $\delta{\mathcal L}_{brane}$. This means that all terms in $ {\mathcal L}^{(1)}_{\Lambda}$ apart from $A_1$ can be absorbed by finite counterterms. A similar conclusion holds for the curvature dependent part of ${\mathcal L}$. The renormalized expression for ${\mathcal L}_{\Lambda}$ becomes
\be
{\mathcal L}_{\Lambda}=-V_{h}^{R}-a^DV_{v}^{R}+\frac{1}{2}(4\pi)^{-D/2}(ka)^D\;A_1(-D/2)\;.\label{3.17}
\ee
The renormalized expression for ${\mathcal L}_R$ becomes
\be
{\mathcal L}_{R}=\Big\lbrack \kappa_{h}^{R}+a^{D-2}\kappa_{v}^{R}+\frac{1}{8\pi}\Big(\frac{1}{6}-\xi\Big)(4\pi)^{1-D/2}(ka)^{D-2}\;A_1(1-D/2) \Big\rbrack R\;.\label{3.18}
\ee
The function $A_1(s)$ was defined in (\ref{A35}) with (\ref{A22}) used for $g_\nu(z)$ in the untwisted field case, and (\ref{B3}) used for twisted fields.

In the next section we look at the possibility of finding self-consistent solutions arising from quantum corrections to the classical theory.

\section{Self-consistent solutions?}
\label{solutions}
\setcounter{equation}{0}

Having obtained the vacuum, or Casimir, energy for the effective $D$-di\-men\-sion\-al theory we can now see if there are any solutions of the Randall-Sundrum type to the quantum corrected field equations. We will argue that the problem is not quite as simple as calling the Casimir energy a potential and looking for its extremum. Rather, we must require that the Einstein field equations hold, and we will see that this has a different requirement.

If $\Gamma$ denotes the full effective action with quantum effects included we must require
\bea
\frac{\delta\Gamma}{\delta g_{\mu\nu}}&=&0\;,\label{4.1}\\
\frac{\delta\Gamma}{\delta a}&=&0\;,\label{4.2}
\eea
with the Randall-Sundrum spacetime (\ref{2.1}) as a solution with $g_{\mu\nu}=\eta_{\mu\nu}$. The classical part of $\Gamma$ comes from (\ref{2.4}) and (\ref{2.5}), and to these contributions we must add the one-loop correction $\Gamma^{(1)}$. If we perform the integration over the extra dimension $y$ using the metric ansatz (\ref{2.1}) with $\sigma$ given by (\ref{2.3}) we obtain
\bea
S_G&=&(16\pi G)^{-1}\int\,dv_x\Big\lbrace \frac{2}{(D-2)k}(1-a^{D-2})R+4kD(1-a^D)\nn
&&-2k(D+1)(1-a^D)-\frac{4\Lambda}{kD}(1-a^D)\Big\rbrace\;.\label{4.3}
\eea
$dv_x=d^Dx|g|^{1/2}$ is the $D$-dimensional invariant volume element. Because we set $g_{\mu\nu}=\eta_{\mu\nu}$ after differentiation with respect to $g_{\mu\nu}$, the term in the curvature $R$ will make no contribution to the field equations. The same will be true for the induced gravity part of the one-loop effective action, as well as for higher order curvature terms that we have not calculated. The induced gravity term is important for relating the overall coefficient of $R$ in the effective action to the physical value of the gravitational constant, which is irrelevant to the field equations in the present case. We are therefore justified in only considering the vacuum energy in the present calculation. Accordingly we will take
\be
\Gamma=S_G-\int dv_x\;F(a)\label{4.4}
\ee
where $F(a)$ includes the brane tension terms from (\ref{2.5}) as well as the Casimir energy.

Variation of $\Gamma$ with respect to $g_{\mu\nu}$ followed by taking $g_{\mu\nu}=\eta_{\mu\nu}$ results in
\be
k(D-1)(1-a^D)-\frac{2\Lambda}{kD}(1-a^D)=8\pi GF(a)\;.\label{4.5}
\ee
Variation of $\Gamma$ with respect to $a$ yields
\be
k(D-1)-\frac{2\Lambda}{kD}=-\frac{8\pi G}{D}\,a^{1-D}\,F'(a)\;,\label{4.6}
\ee
after setting $g_{\mu\nu}=\eta_{\mu\nu}$. Combining (\ref{4.5}) and (\ref{4.6}) gives us a requirement for $F(a)$~:
\be
0=(1-a^D)a F'(a)+D a^D F(a)\;.\label{4.7}
\ee

As a check on this last result, if we set
\be
F(a)=V_v a^D+V_h\;,\label{4.8}
\ee
corresponding to the brane tension terms in (\ref{2.5}), 
we find the condition
\be
V_v+V_h=0\;.\label{4.9}
\ee
This reproduces the result on the balancing of the two brane 
tensions needed in the Randall-Sundrum model. The remaining
requirements (\ref{2.7}) and (\ref{2.8})  follow upon using 
(\ref{4.9}) in (\ref{4.5}) and (\ref{4.6}).

If we now include the vacuum energy by taking
\be
F(a)=V_va^D+V_h+f(a)\;,\label{4.10}
\ee
in place of the classical expression (\ref{4.8}), we must have
\be
0= D a^D (V_v + V_h) + (1-a^{D})af'(a)+Da^D f(a)\label{4.11}
\ee
holding. 
This is clearly not just a simple minimization condition 
on the potential. From (\ref{3.17}) we have
\be
f(a)=-\frac{1}{2}(4\pi)^{-D/2}k^Da^DA_1(-D/2)\label{4.12}
\ee
with $A_1$ defined in (\ref{A35}). If we take
\be
I(a)=A_1(-D/2)\;,\label{4.13}
\ee
then we must have 
\be
0= {2 D (4 \pi)^{D/2} \over k^D}(V_v + V_h) - DI(a)
- a(1-a^D)I'(a)\;.\label{4.14}
\ee
This condition must be fulfilled if the Randall-Sundrum 
spacetime is to be a self-consistent solution to the 
effective 4-dimensional Einstein equations with quantum 
effects included.

In order to satisfy the self-consistency relation (\ref{4.14}), we can use the freedom 
to tune the brane tensions. A possible choice is to require that the 
balancing condition (\ref{4.9}) between the brane tensions still holds. For this choice it is easy 
to see that there are no solutions to (\ref{4.14}). To see 
this is quite simple. It is easy to show that $I(a)$ is negative for 
$0<a<1$, and is monotonically decreasing. The right hand side of 
(\ref{4.14}) can never vanish. However, there is no need to require that $V_v + V_h =0$ 
when quantum effects are included\footnote{We would like to thank J. Garriga 
for having pointed this out to us.}, and relaxing this condition allows 
solutions to (\ref{4.14}). The interesting thing is that it is possible to find 
self-consistent solutions for values of $a$ which solve the hierarchy problem only 
at the price of severe fine tuning of the brane tensions\footnote{This has 
been noted in \cite{Garriga} for the case of massless minimally coupled and 
conformally coupled fields}.
This is a more general result than 
\cite{GR} who only were interested in self-consistent solutions which had 
values for $a$ consistent with the solution to the gauge hierarchy problem 
({\em ie\/} $a<<1$), and who only examined the Casimir, or vacuum, energy. 

\section{Massless, conformally coupled fields}
\label{conformal}
\setcounter{equation}{0}

The case of massless, conformally coupled scalars deserves special mention. Taking $m^2=0$ and $\xi=(D-1)/(4D)$  in the order of the Bessel functions determined in (\ref{2.18}) gives $\nu=1/2$. Because Bessel functions of half-integer order can be given in terms of elementary functions, the analysis becomes much simpler than for general $\nu$. 
If we examine the twisted fields first, and use
\bea
J_{1/2}(z)&=&\sqrt{\frac{2}{\pi z}}\;\sin z\;,\label{con1}\\
Y_{1/2}(z)&=&-\sqrt{\frac{2}{\pi z}}\;\cos z\;,\label{con2}
\eea
it is easy to see that $\tilde{F}_{1/2}(z)$ defined in (\ref{2.20d}) is given by
\be
\tilde{F}_{1/2}(z)=-\frac{4\sin(1-a)z}{\pi^2\sqrt{a}\;z}\;.\label{con3}
\ee
We therefore have the mass eigenvalues given by
\be
m_n=\frac{\pi k a}{(1-a)}\;n\;,\ {\rm for}\ n=1,2,3,\ldots\;.\label{con4}
\ee
We can evaluate the sum over modes in ${\mathcal L}^{(1)}_{\Lambda}$ directly using the Riemann $\zeta$-function, without using the contour integral method of Appendix~B. It is easy to see that
\bea
{\mathcal L}^{(1)}_{\Lambda}&=&\frac{1}{2}(4\pi)^{-D/2}\left(\frac{\pi k a}{1-a}\right)^D\ell^\epsilon\;\Gamma(-D/2)\zeta(-D)\nn\\
&=&\frac{1}{2\sqrt{\pi}}(4\pi)^{-D/2}\left(\frac{ k a}{1-a}\right)^D \Gamma\left(\frac{D+1}{2}\right) \zeta(D+1)\;.\label{con5}
\eea
In the last line we have let $\epsilon\rightarrow0$, since the result is analytic at $\epsilon=0$, and used some of the relations of the $\Gamma$- and $\zeta$-functions to simplify the result. Although no infinte renormalizations are required, it is still necessary to do finite ones.

As a check on this result, we can now apply the method of Appendix~B. We need to know $A_1(-D/2)$ which is given by 
\bea
A_1(-D/2)&=&\frac{1}{\Gamma(1+D/2)}\int\limits_0^\infty dz\,z^D\frac{d}{dz}\ln\tilde{g}_{1/2}(z)\nn\\
&=& -\frac{2}{\Gamma(D/2)}\int\limits_0^\infty dz\,z^{D-1}\ln\left\lbrack \frac{1-e^{-2(1-a)z}}{1-e^{-2z}}\right\rbrack\;.\label{con6}
\eea
(In the second line we have integrated by parts and made use of the Bessel functions (\ref{con1},\ref{con2}) for $\nu=1/2$. ) The integral is easily evaluated in terms of the Riemann $\zeta$-function with the result
\be
A_1(-D/2)=\frac{1}{\sqrt{\pi}}\,\Gamma\left(\frac{D+1}{2}\right)\zeta(D+1)\big\lbrack(1-a)^{-D}-1\big\rbrack                                                                                                       \;.\label{con7}
\ee
Note that when expanded in powers of $a$, $A_1$ begins with order $a$ as found in the case of general $\nu$. The result in (\ref{con7}) gives rise to the same result as found by a direct summation of the mass eigenvalues, after imposing the same renormalization conditions, and provides a useful check on the method. 

The untwisted conformally coupled scalar field case was also dealt with in Ref.~\cite{Garriga} who exploited the conformal invariance to relate the problem to one in flat spacetime with boundaries present. It is worth emphasizing that this requires the theory to have no conformal anomaly, or else extra terms will arise when transforming back to the original spacetime. The fact that no such terms do arise is guaranteed by our method which makes no use of conformal transformations. In effect we have shown that there is no conformal anomaly for conformally coupled scalars in the Randall-Sundrum spacetime in any dimension. Vanishing of the conformal anomaly is guaranteed by the vanishing of $\tilde{d}_k$ when $\nu=1/2$. (This is easily checked to be true for the coefficients listed in (\ref{B5a}--\ref{B5f}).)

We now look at the case of untwisted fields. The functions $j_\nu$ and $y_\nu$ defined in (\ref{2.16},\ref{2.17}) are easily seen to be
\bea
j_{1/2}(z)&=&\sqrt{\frac{2z}{\pi}}\;\cos z\;,\label{con8}\\
y_{1/2}(z)&=&\sqrt{\frac{2z}{\pi}}\;\sin z\;,\label{con9}
\eea
We then find 
\be
F_{1/2}(z)=-\frac{4\sqrt{a}}{\pi^2}\;z\,\sin(1-a)z\;,\label{con10}
\ee
where $F_{1/2}(z)$ was defined in (\ref{2.15}). The mass eigenvalues are given by the same expression (\ref{con4}) as we had in the twisted field case. A direct summation over the mass eigenvalues again leads to (\ref{con5}).

We can also check the result by using the contour integral method of Appendix~A. The integral $A_1(-D/2)$ becomes
\be
A_1(-D/2)=\frac{1}{\sqrt{\pi}}\,\Gamma\left(\frac{D+1}{2}\right)\zeta(D+1)\big\lbrack(1-a)^{-D}+1- 
2^{-D}\big\rbrack \;.\label{con11}
\ee
The only difference with the twisted field result is the constant term. The $(1-a)^{-D}$ dependence is the same. Note that as a consequence, when $a$ is small, $A_1$ is approximately a constant, rather than proportional to $a$. This would seem to contradict the general results established in Appendix~A. The root cause of this lies in the behaviour of $F_{1/2}(z)$ as $z\rightarrow0$. Clearly $F_{1/2}(z)\sim z^2$ as $z\rightarrow0$, which does not satisfy the assumption we made in Appendix~A that $F_{1/2}(0)$ is finite and non-zero. In this particular case we must remove the zero of $F_{1/2}(z)$ at $z=0$ and take instead $z^{-2}F_{1/2}(z)$ which does satisfy the conditions of being finite and non-zero as $z\rightarrow0$ allowing the deformation of contour used in Appendix~A. If we compare $z^{-2}F_{1/2}(z)$ with $\tilde{F}_{1/2}(z)$ it is obvious that the same result will now be obtained for $A_1(-D/2)$ for the untwisted field as we found for the twisted field, and which is in agreement with the direct summation over mass modes. In any case, the difference between the twisted and untwisted fields for $A_1$ results in a term proportional to $a^D$ in the effective action which can be dealt with by renormalization. 

For either twisted or untwisted field configurations it is easy to show 
that the only solutions to (\ref{4.14}) is $a=0$ for $V_v+V_h=0$. 
When this last condition is relaxed, it is possible to find self-consistent solutions, 
but if we demand the hierarchy problem to be solved a severe fine tuning of the brane 
tensions is required.

\section{Discussion and conclusions}

In the present paper we have examined the one-loop radiative corrections to the effective action arising from the quantization of a massless scalar field non-minimally coupled to the curvature of the background spacetime proposed by Randall and Sundrum \cite{RS}.  Although it is not possible to obtain an exact result for general, curved versions of the Randall-Sundrum spacetime, it is possible to obtain an expansion in powers of the curvature.  The vacuum (Casimir) energy is the first term in this expansion.  The vacuum energy has received attention \cite {Garriga,DJTPLB,GR} due to its possible role in stabilising the radius of the extra dimension in a self-consistent way, as an alternative to the mechanism proposed in \cite{GW2}.  Because of the different conclusions and inconsistent results obtained in previous analyses, we have presented a careful and complete calculation here.  Differences and comparisons of our results are made with Refs.~\cite{Garriga,DJTPLB,GR} at appropriate stages.

One important clarification concerns the boundary conditions adopted for the scalar field.  We emphasized that there 
were two inequivalent field configurations (conventionally referred to as twisted and twisted \cite{Isham}).  
Either can be used, and in general the results will be different.  

Another clarification concerns the renormalisation of the theory.  We presented a full calculation for both field configurations for the generally coupled massive scalar field with all relevant counterterms indicated.  To do this it is necessary to perform a full analytic continuation of relevant integrals as outlined in Appendices~A and B.  As we emphasized, the same method can be used for both twisted and untwisted fields, but the precise details of the analytic continuation are slightly different due to the different asymptotic behaviour of the Bessel functions between the two field configurations.  Ref.~\cite{GR} did not perform the detailed analytic continuation, and really only analyzed the untwisted field configuration.  Nevertheless, as we showed above, the essential conclusions of their paper concerning the renormalisation are correct in either case.  One reason for presenting the detailed calculation of the counterterms is that the pole term in dimensional regularization is related to the conformal anomaly for the scalar field.

As noted in the older Kaluza-Klein models based on spaces with homogeneous extra dimensions \cite{CW}, a full analysis of self-consistency can only be achieved if the quantum correction to the gravitational part of the effective action is known.  We presented the necessary analysis for this above.  In order to renormalize the theory when the branes were curved, it was necessary to add on additional contributions to the brane tensions involving the curvature.  This suggest that the brane sector of the model has to be viewed as a series, with the brane tensions as the first term in the expansion.

Another clarification concerns the conditions for self-consistency of the 
Randall-Sundrum spacetime.  With flat branes we showed that the induced 
gravity contribution was not important, in contrast to the situation in 
\cite{CW}.  Additionally, it was argued that the requirement of 
self-consistency is not obtained simply by looking at the minimum of the 
Casimir energy.  By looking at the quantum corrected Einstein equations 
for the effective 4-dimensional (or more generally $D$-dimensional) theory 
we obtained a consistency condition which the radius of the extra 
dimension had to satisfy.  We found no solutions to this if we demand 
$V_v+V_h=0$, meaning that it is not possible to stabilize the radius by quantum
effects coming from bulk scalar fields in this case. This conclusion can
be altered by relaxing the balancing condition $V_v+V_h=0$, even if we
demand that the gauge hierarchy problem also be solved, but only at the price 
of fine tuning the brane tensions.

We also examined massless conformally coupled scalars fields as a limiting case of our general results.  This provides a useful check on the results, since it is possible to perform the calculations in a different, and much simpler, way.  It also provides a comparison with Ref.~\cite{Garriga} where a conformal transformation was used to relate the calculation to one in a flat spacetime with boundaries.  A curious feature of the massless conformally coupled case is that the imposition of twisted or untwisted boundary conditions makes no difference to the final result.  As we mentioned above, our calculation provides a proof that there is no conformal anomaly for the Randall-Sundrum spacetime in any number of dimensions.

An interesting point concerns the relation between viewing the quantized scalar field as an effective 4-dimensional theory, and a 5-dimensional theory.  From the 5-dimensional viewpoint, self-consistency must be satisfied from the 5-dimensional Einstein field equations.  To do this, it is necessary to evaluate not just the total energy, but rather the energy density. (More specifically we must know the vacuum expectation value of the stress-energy tensor.) Due to the boundary conditions and the nature of the spacetime, the vacuum energy density for the 5-dimensional scalar field will not be expected to be constant in general.  It would be interesting to study this problem, although the presence of Bessel functions in the mode eigenfunctions would make this rather difficult.

Final comments concern the whole question of using quantum effects to 
obtain a self-consistent result for the radius of the extra dimensions.  
The conclusions of Ref.~\cite{GR} were rather negative concerning the 
possibility of simultaneously solving the gauge hierarchy problem and 
self-consistency.  It is still interesting to ask if it is possible to 
obtain self-consistent solutions at all, regardless of the size of the 
extra dimensions.  All that we currently know about this is that it is 
possible to do so with bulk scalar fields, but, fine tuning of the brane tensions 
is required when we demand that the hierarchy problem is simultaneously solved.

Whether or not a theory with more realistic matter content can do so is 
an open question.  Finally, even if it is possible to find a self-consistent 
solution there still remains the problem of determining if it is stable.

\section*{Acknowledgements}

A. Flachi is grateful to the University of Newcastle upon Tyne for the award of a Ridley Studentship. We are grateful to J. Garriga for comments concerning Ref.~\cite{Garriga}, and to I.~G.~Moss for discussions on the conformally coupled case.

\appendix\section{Evaluation of the sums for untwisted fields}
\setcounter{equation}{0}
\renewcommand{\theequation}{\mbox{A.\arabic{equation}}}

\begin{figure}[htb]
\begin{center}
\leavevmode
\epsfxsize=100mm
\epsffile{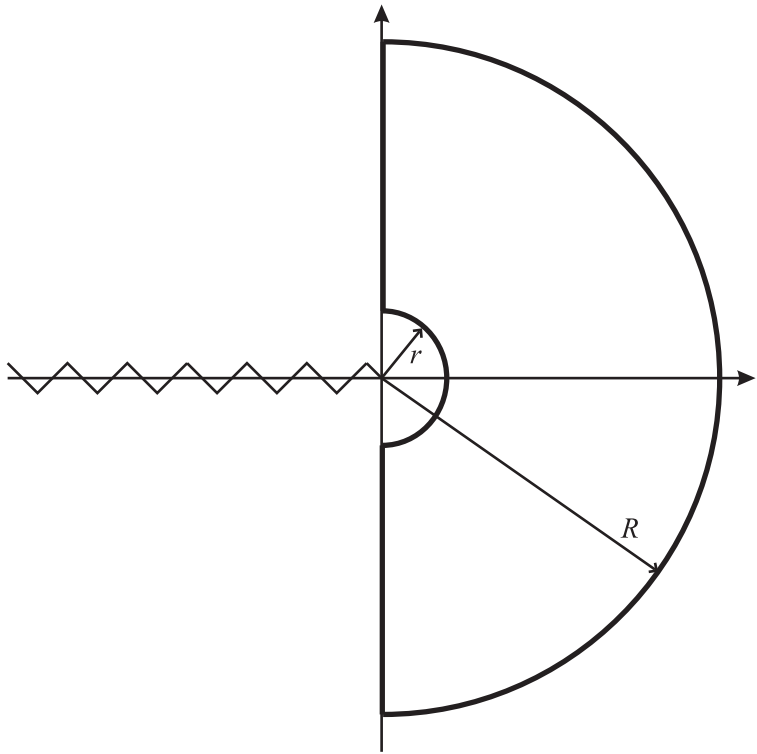}
\end{center}
\caption{\footnotesize This shows the contour ${\mathcal C}$ used in the evaluation of (\ref{A3}). The contour is traversed in the counterclockwise direction and the limits $r\rightarrow0$ and $R\rightarrow\infty$ are understood. There is a branch cut in the integrand that we choose to lie along the negative real axis.}\label{contour}
\end{figure}

In this appendix we outline the evaluation of the sum over modes needed to calculate the vacuum energy and induced gravity terms. Begin by defining
\be
\omega(s)=\Gamma(s)\sum_{n=1}^{\infty}x_n^{-2s}\;,\label{A1}
\ee
where $x_n$ is the $n^{\rm th}$ positive solution to $F_\nu(x_n)=0$ with $F_\nu(z)$ defined in terms of Bessel functions in (\ref{2.15}). For large $n$ we know $x_n\sim n$ so that (\ref{A1}) is defined initially for $\Re(s)>1/2$ where the sum converges. We want to analytically continue $\omega(s)$ to $s=-D/2$ and $1-D/2$. The basic method for doing this is to convert the sum into a contour integral. A simple application of the residue theorem shows that
\be
\omega(s)=\frac{\Gamma(s)}{2\pi i}\int_{\mathcal C}dz\,z^{-2s}\,\frac{d}{dz}\ln\,F_\nu(z)\;,\label{A3}
\ee
where ${\mathcal C}$ is any contour which encloses the positive zeros of $F_\nu(z)$ defined in (\ref{2.15}). A convenient choice is shown in Fig.~\ref{contour}.

This procedure is basically the same as that used in Refs.~\cite{Les,Bordag} for studying the Casimir effect in a spherical shell. Ref.~\cite{Les} uses an integration by parts to remove the derivative in (\ref{A3}), and it is this basic starting point that Refs.~\cite{Garriga,GR} follow. It is more convenient to stick with (\ref{A3}) for our purposes, although either method could be used. A difference between our approach and that of Ref.~\cite{Bordag} is that we do not need to use the uniform asymptotic expansion of the Bessel functions, making our calculation slightly simpler.

By using the behaviour of the Bessel functions for small $z$ it can be shown that the contribution from the semicircle of radius $r$ behaves like $r^{2-2s}$ for small $r$. This means that if we restrict ourselves to the strip $\frac{1}{2}<\Re(s)<1$ the contribution to the contour coming from this portion of the contour will vanish as $r\rightarrow0$. (The fact that $F_\nu(0)$ is finite and non-zero is crucial to this result.) For the semicircle of radius $R$, the integrand behaves like $R^{-2s}$ for large $R$, which follows from using the large argument expansion for the Bessel functions in $F_\nu(z)$. The contribution from this portion of the contour also vanishes if we let $R\rightarrow\infty$ and restrict ourselves to the strip $\frac{1}{2}<\Re(s)<1$. The only contributions to the integral for $\frac{1}{2}<\Re(s)<1$ come from the portions of the contour along the positive and negative imaginary axis. After a short calculation it can be shown that
\be
\omega(s)=\frac{1}{\Gamma(1-s)}\int\limits_0^\infty {dz}\,z^{-2s}\frac{d}{dz}\ln\,P_\nu(z)\;,\label{A4}
\ee
with
\be
P_\nu(z)=\frac{2}{\pi}\left\lbrack i_\nu(az)k_\nu(z)-i_\nu(z)k_\nu(az)\right\rbrack\;.\label{A5}
\ee
and
\bea
i_\nu(z)&=&\frac{D}{2}(1-4\xi)I_\nu(z)+zI^{\prime}_{\nu}(z)\;,\label{A6}\\
k_\nu(z)&=&\frac{D}{2}(1-4\xi)K_\nu(z)+zK^{\prime}_{\nu}(z)\;.\label{A7}
\eea

We now wish to analytically continue the expression (\ref{A4}) for $\omega(s)$ out of the strip $\frac{1}{2}<\Re(s)<1$ to $\Re(s)<\frac{1}{2}$. The simplest way to do this is to note that the impediment to convergence of (\ref{A4}) for $\Re(s)<\frac{1}{2}$ comes from the behaviour of $P_\nu(z)$ at large $z$. We will define
\bea
I_\nu(z)&=&\frac{e^z}{\sqrt{2\pi z}}\,\Sigma^{(I)}(z)\;,\label{A8}\\
K_\nu(z)&=&\sqrt{\frac{\pi}{2 z}}\,e^{-z}\,\Sigma^{(K)}(z)\;,\label{A9}
\eea
which gives
\bea
i_\nu(z)&=&\sqrt{\frac{z}{2\pi}}\,e^z\,\Sigma_\nu^{(I)}(z)\;,\label{A10}\\
k_\nu(z)&=&-\sqrt{\frac{\pi z}{2}}\,e^{-z}\,\Sigma_\nu^{(K)}(z)\;,\label{A11}
\eea
from (\ref{A6}) and (\ref{A7}) where
\bea
\Sigma_\nu^{(I)}(z)&=&\Sigma^{(I)}(z)+\frac{d}{dz}\Sigma^{(I)}(z)+\frac{1}{2z}\lbrack D(1-4\xi)-1\rbrack \Sigma^{(I)}(z)\;,\label{A12}\\
\Sigma_\nu^{(K)}(z)&=&\Sigma^{(K)}(z)-\frac{d}{dz}\Sigma^{(K)}(z)-\frac{1}{2z}\lbrack D(1-4\xi)-1\rbrack \Sigma^{(K)}(z)\;.\label{A13}
\eea
For large $z$ the asymptotic expansions for the Bessel functions show that
\be
\Sigma^{(I)}(z)\simeq\sum_{k=0}^{\infty}\alpha_k\,z^{-k}\;,\label{A14}
\ee
where
\be
\alpha_k=\frac{(-1)^k\,\Gamma(\nu+k+\frac{1}{2})}{2^k\,k!\,\Gamma(\nu-k+\frac{1}{2})}\;,\label{A15}
\ee
and
\be
\Sigma^{(K)}(z)\simeq\Sigma^{(I)}(-z)\;.\label{A16}
\ee

Because the divergent behaviour of $\omega(s)$ comes from the large $z$ behaviour of the integrand, we will split up
\be
\omega(s)=\omega_1(s)+\omega_2(s)\;,\label{A17}
\ee
where
\bea
\omega_1(s)&=&\frac{1}{\Gamma(1-s)}\int\limits_0^1 {dz}\,z^{-2s}\frac{d}{dz}\ln\,P_\nu(z)\;,\label{A18}\\
\omega_2(s)&=&\frac{1}{\Gamma(1-s)}\int\limits_1^\infty {dz}\,z^{-2s}\frac{d}{dz}\ln\,P_\nu(z)\;.\label{A19}
\eea
(The choice of 1 as the upper limit in (\ref{A18}) is not important; any value greater than 0 will be equally good.) Analysis of the terms in $P_\nu(z)$ responsible for divergence when $\Re(s)<\frac{1}{2}$ leads us to define
\be
P_\nu(z)=\frac{1}{\pi}a^{1/2}ze^{(1-a)z}\,\Sigma_\nu^{(I)}(z)\Sigma_\nu^{(K)}(az)g_\nu(z)\;,\label{A20}
\ee
where
\bea
g_\nu(z)&=&1-e^{-2(1-a)z}\,\frac{\Sigma_\nu^{(K)}(z)\Sigma_\nu^{(I)}(az)}{\Sigma_\nu^{(I)}(z)\Sigma_\nu^{(K)}(az)}\label{A21}\\
&=&1-\frac{k_\nu(z)i_\nu(az)}{i_\nu(z)k_\nu(az)}\;.\label{A22}
\eea
Substitution of (\ref{A20}) into (\ref{A18}) results in
\bea
\omega_1(s)&=&\frac{1}{\Gamma(1-s)}\left\lbrace-\frac{1}{2s}+\frac{1-a}{1-2s}\right\rbrace+\frac{1}{\Gamma(1-s)}\int\limits_0^1 {dz}\,z^{-2s}\Big\lbrace\frac{d}{dz}\ln\,g_\nu(z)\nn
&&+\frac{d}{dz}\ln\Big\lbrack\Sigma_\nu^{(I)}(z)\Sigma_\nu^{(K)}(az)\Big\rbrack\Big\rbrace\label{A23}
\eea
where the result has been analytically continued to $\Re(s)<0$. Substitution of (\ref{A20}) into (\ref{A19}), assuming initially $\frac{1}{2}<\Re(s)<1$, and then analytically continuing to $\Re(s)<0$, results in
\bea
\omega(s)&=&\frac{1}{\Gamma(1-s)}\int\limits_0^\infty {dz}\,z^{-2s}\,\frac{d}{dz}\ln\,g_\nu(z) \nn
&&+ \frac{1}{\Gamma(1-s)}\int\limits_0^1 {dz}\,z^{-2s} \frac{d}{dz}\ln\Big\lbrack\Sigma_\nu^{(I)}(z)\Sigma_\nu^{(K)}(az)\Big\rbrack\nn
&&+\frac{1}{\Gamma(1-s)}\int\limits_1^\infty {dz}\,z^{-2s} \frac{d}{dz}\ln\Big\lbrack\Sigma_\nu^{(I)}(z)\Sigma_\nu^{(K)}(az)\Big\rbrack\;.\label{A24}
\eea
The first two terms are analytic for $\Re(s)<0$; however, the last term has poles at various values of $s$ which we need to determine if we are to perform the required analytic continuation. 

We have constructed $\Sigma_\nu^{(I)}(z)$ so that $\Sigma_\nu^{(I)}(z)\sim1+\mathcal{O}(z^{-1})$ as $z\rightarrow\infty$. This means that we can write
\be
\ln\,\Sigma_\nu^{(I)}(z)\simeq\sum_{k=1}^{\infty}d_k\,z^{-k}\label{A25}
\ee
for large $z$, with some coefficients $d_k$. The coefficients $d_k$ will determine the poles of $\omega(s)$. Because of (\ref{A16}) we have
\be
\ln\Big\lbrack\Sigma_\nu^{(I)}(z)\Sigma_\nu^{(K)}(az)\Big\rbrack\simeq\sum_{k=1}^{\infty}\sigma_k\,z^{-k}\label{A26}
\ee
with
\be
\sigma_k=\Big\lbrack1+\frac{(-1)^k}{a^k}\Big\rbrack\,d_k\;.\label{A27}
\ee
Using (\ref{A14}) and (\ref{A15}) in (\ref{A25}) it is easy to calculate the coefficients $d_k$. The first four coefficients are found to be
\bea
d_1&=&\alpha-\frac{3}{8}-\frac{1}{2}\nu^2\;,\label{A28a}\\
d_2&=&\frac{1}{4}\nu^2-\frac{3}{16}+\frac{1}{2}\alpha-\frac{1}{2}\alpha^2\;,\label{A28b}\\
d_3&=&-\frac{1}{2}\alpha\nu^2-\frac{1}{2}\alpha^2+\frac{3}{8}\alpha+\frac{23}{48}\nu^2+\frac{1}{24}\nu^4-\frac{21}{128}+\frac{1}{3}\alpha^3\;,\label{A28c}\\
d_4&=&-\alpha\nu^2-\frac{1}{2}\alpha^2+\frac{3}{8}\alpha+\frac{13}{16}\nu^2-\frac{1}{8}\nu^4\nn
&&+\frac{1}{2}\alpha^3+\frac{1}{2}\alpha^2\nu^2-\frac{1}{4}\alpha^4-\frac{27}{128}\;,\label{A28d}
\eea
where
\be
\alpha=\frac{D}{2}(1-4\xi)\;.\label{A29}
\ee
(For $\alpha=2$, which is the case for $D=4$ and $\xi=0$,  the results reduce to those given in \cite{DJTPLB}.) By adding and subtracting terms to the last integrand of (\ref{A24}) we may analytically continue the result to $\Re(s)<0$. The more negative $\Re(s)$ becomes, the more terms we need to subtract. In this way we obtain
\bea
\omega(s)&=&\frac{1}{\Gamma(1-s)}\int\limits_0^\infty {dz}\,z^{-2s}\,\frac{d}{dz}\ln\,g_\nu(z) \nn
&&+ \frac{1}{\Gamma(1-s)}\int\limits_0^1 {dz}\,z^{-2s} \frac{d}{dz}\ln\Big\lbrack\Sigma_\nu^{(I)}(z)\Sigma_\nu^{(K)}(az)\Big\rbrack\nn
&&+\frac{1}{\Gamma(1-s)}\int\limits_1^\infty {dz}\,z^{-2s} \frac{d}{dz}\left\lbrace\ln\Big\lbrack\Sigma_\nu^{(I)}(z)\Sigma_\nu^{(K)}(az)\Big\rbrack-\sum_{k=1}^{N}\sigma_kz^{-k}\right\rbrace\nn
&&-\frac{1}{\Gamma(1-s)}\sum_{k=1}^{N}\frac{k\sigma_k}{(k+2s)}\;.\label{A30}
\eea
Provided that $N$ is chosen so that $\Re(N+1+2s)>0$ the analytic continuation to any value of $s$ with $\Re(s)<0$ may be performed. All of the possible poles can be found in the last term of (\ref{A30}). Some procedure like the one we have described is essential to provide the proper analytic continuation of $\omega(s)$ to $\Re(s)<0$.

We are interested in the value of $\omega(s)$ at $s=p-D/2$ for $p=0,1$. There is a simple pole at this value of $s$ coming from the $k=D-2p$ term in the sum in the last term of (\ref{A30}). We can write
\be
\omega(s)=\omega_{\rm pole}(s)+\omega_{\rm reg}(s)\label{A31}
\ee
where
\be
\omega_{\rm pole}(s)=-\frac{1}{\Gamma(1-s)}\frac{(D-2p)\sigma_{D-2p}}{(D-2p+2s)}\label{A32}
\ee
contains the pole term (as well as a finite term found by expanding about the pole), and $\omega_{\rm reg}(s)$ is regular ({\em ie}\/ analytic) at $s=p-D/2$ and is given by an expression like (\ref{A30}) but with the term $k=D-2p$ omitted from the sum in the last term. We are free to simply set $s=p-D/2$ in $\omega_{\rm reg}(s)$, where $D$ is the physical dimension ({\em ie}\/ 4 for the Randall-Sundrum model). 

The most important thing about the regular part $\omega_{\rm reg}(s)$ is the dependence on $a$. By splitting up the logarithms in (\ref{A30}) it can be seen that
\be
\omega_{\rm reg}(s)=C(s)+A_1(s)+A_2(s)\label{A33}
\ee
where
\bea
C(s)&=&\frac{1}{\Gamma(1-s)}\int\limits_0^1 {dz}\,z^{-2s} \frac{d}{dz}\ln\Sigma_\nu^{(I)}(z)\nn
&&+\frac{1}{\Gamma(1-s)}\int\limits_1^\infty {dz}\,z^{-2s} \frac{d}{dz}\left\lbrace\ln\Sigma_\nu^{(I)}(z)-\sum_{k=1}^{N}d_kz^{-k}\right\rbrace\nn
&&-\frac{1}{\Gamma(1-s)}{\sum_{k=1}^{N}}'\frac{kd_k}{(k+2s)}\label{A34}
\eea
is independent of $a$, and 
\bea
A_1(s)&=&\frac{1}{\Gamma(1-s)}\int\limits_0^\infty {dz}\,z^{-2s}\,\frac{d}{dz}\ln\,g_\nu(z) \label{A35}\\
A_2(s)&=&\frac{1}{\Gamma(1-s)}\int\limits_0^1 {dz}\,z^{-2s} \frac{d}{dz}\ln\Sigma_\nu^{(K)}(az)\nn
&&+\frac{1}{\Gamma(1-s)}\int\limits_1^\infty {dz}\,z^{-2s} \frac{d}{dz}\left\lbrace\ln\Sigma_\nu^{(K)}(az)-\sum_{k=1}^{N}\frac{d_k(-1)^k}{(az)^{k}}\right\rbrace\nn
&&-\frac{1}{\Gamma(1-s)}{\sum_{k=1}^{N}}'\frac{k(-1)^kd_k}{(k+2s)a^k}\label{A36}
\eea
contain the $a$-dependence. The prime on the sums in (\ref{A34}) and (\ref{A36}) denote that the term $k=D-2p$, which gives rise to the pole, is omitted. 

The $a$-dependence of (\ref{A36}) can be made more explicit by a change of variable $z\rightarrow a^{-1}z$ in both integrals~:
\bea
A_2(s)&=&\frac{a^{2s}}{\Gamma(1-s)}\int\limits_0^a {dz}\,z^{-2s} \frac{d}{dz}\ln\Sigma_\nu^{(K)}(z)\nn
&&+\frac{a^{2s}}{\Gamma(1-s)}\int\limits_a^\infty {dz}\,z^{-2s} \frac{d}{dz}\left\lbrace\ln\Sigma_\nu^{(K)}(z)-\sum_{k=1}^{N}\frac{d_k(-1)^k}{z^{k}}\right\rbrace\nn
&&-\frac{1}{\Gamma(1-s)}{\sum_{k=1}^{N}}'\frac{k(-1)^kd_k}{(k+2s)a^k}\;.\label{A37}
\eea
Now split up the integration range of the second integral as $\displaystyle{\int\limits_a^\infty=\int\limits_a^1 +\int\limits_1^\infty}$. After a short calculation we are left with
\bea
A_2(s)&=&a^{2s}A(s)-\frac{a^{2s}}{\Gamma(1-s)}\int\limits_a^1\,dz\,z^{-2s}\frac{d}{dz}\sum_{k=1}^{N}\frac{(-1)^kd_k}{z^k}\nn
&& -\frac{1}{\Gamma(1-s)}{\sum_{k=1}^{N}}'\frac{k(-1)^kd_k}{(k+2s)a^k}\label{A38}
\eea
where
\bea
A(s)&=&\frac{1}{\Gamma(1-s)}\int\limits_0^1 {dz}\,z^{-2s} \frac{d}{dz}\ln\Sigma_\nu^{(K)}(z)\nn
&&+\frac{1}{\Gamma(1-s)}\int\limits_1^\infty {dz}\,z^{-2s} \frac{d}{dz}\left\lbrace\ln\Sigma_\nu^{(K)}(z)-\sum_{k=1}^{N}\frac{d_k(-1)^k}{z^{k}}\right\rbrace\label{A39}
\eea
is independent of $a$ (and is analytic for $\Re(s)<0$). Finally we may perform the integration in (\ref{A38}) to find that
\be
A_2(s)=a^{2s}\bar{A}(s)-\frac{(D-2p)(-1)^D}{\Gamma(1+D/2-p)}\,d_{D-2p}\,a^{2p-D}\ln a\label{A40}
\ee
with
\be
\bar{A}(s)=A(s)-\frac{1}{\Gamma(1-s)}{\sum_{k=1}^{N}}'\frac{k(-1)^kd_k}{(k+2s)}\;.\label{A41}
\ee

We can conclude that for $s=p-D/2$
\bea
\omega_{\rm reg}(p-D/2)&=&C(p-D/2)+A_1(p-D/2)+a^{2p-D}\bar{A}(p-D/2)\nn
&&-\frac{1}{\Gamma(1-s)}{\sum_{k=1}^{N}}'\frac{k(-1)^kd_k}{(k+2s)}\label{A42}
\eea
where $C(s),\bar{A}(s)$ are independent of $a$.

Now turn to a similar analysis for the simpler pole part of $\omega(s)$ given in (\ref{A32}). Because there is  a pole, we will use dimensional regularisation and set $D\rightarrow D+\epsilon$, with $s=p-(D+\epsilon)/2$. Note that it is only in the $s$-dependent terms that we replace $D\rightarrow D+\epsilon$, since the factor of $(D-2p)\sigma_{D-2p}$ came from a term in the asymptotic expansion which involved only integral powers. Because $\sigma_k\propto d_k$, and $d_k$ depends on the order of the Bessel functions $\nu$, there can be a $D$-dependence hidden inside $\sigma_{D-2p}$. However it is easy to see that this dependence can only affect terms which are constant, or proportional to $a^{2p-D}$. Such terms can be absorbed into redefined functions $C(p-D/2)$ and $\bar{A}(p-D/2)$ respectively, which we show in Sec.~\ref{renormalization} can be removed by finite renormalizations. There is no point in giving the explicit form of such terms, and we can safely ignore the $\epsilon$-dependence of $\sigma_{D-2p}$. It is possible to write
\be
\omega_{\rm pole}(s)=\frac{\alpha_{-1}}{\epsilon}+\alpha_0+\alpha_1\,a^{2p-D}\label{A43}
\ee
for
\be
\alpha_{-1}=\frac{(D-2p)}{\Gamma(1+D/2-p)}\left\lbrack 1+\frac{(-1)^D}{a^{D-2p}}\right\rbrack d_{D-2p}\label{A44}
\ee
with $\alpha_0$ and $\alpha_1$ independent of $a$. (As explained above, it is a straightforward matter to obtain explicit results for $\alpha_0$ and $\alpha_1$ at $s=p-D/2$ by expanding (\ref{A32}) about the pole.)

\section{Evaluation of the sums for twisted fields}
\setcounter{equation}{0}
\renewcommand{\theequation}{\mbox{B.\arabic{equation}}}

In this appendix we look at the calculation of $\omega(s)$ using the solution to (\ref{2.20c}). The details are very similar to those of the preceding appendix, so we will be brief. $\omega(s)$ is expressed as a contour integral as in (\ref{A3}) where $\tilde{F}_\nu(z)$ appears in place of $F_\nu(z)$.  (\ref{A4}) still holds, but with $P_\nu(z)$ now given by
\be
P_\nu(z)=\frac{2}{\pi}\left\lbrack I_\nu(z)K_\nu(az)-I_\nu(az)K_\nu(z)\right\rbrack\;.\label{B1}
\ee
Using (\ref{A8}) and (\ref{A9}) results in
\be
P_\nu(z)=\frac{1}{\pi}a^{-1/2}z^{-1}e^{(1-a)z}\Sigma^{(I)}(z)\Sigma^{(K)}(az)g_\nu(z)\label{B2}
\ee
where
\be
g_\nu(z)=1-\frac{K_\nu(z)I_\nu(az)}{K_\nu(az)I_\nu(z)}\;.\label{B3}
\ee
The difference between the twisted and untwisted cases lies in the asymptotic behaviour of the integrand of the contour integral for large $|z|$. (Compare (\ref{A20}) with (\ref{B2}).) In the untwisted case the leading term in the asymptotic expansion comes from the derivative of the Bessel functions in (\ref{A6},\ref{A7}). These terms are absent from the twisted field expressions. Because of this different asymptotic behaviour, which lead to divergent integrals dealt with by adding and subtracting terms as discussed in the previous appendix, it is difficult to deal with both twisted and untwisted fields in one single calculation.

The steps in the analytic continuation of $\omega(s)$ follow much as before with slightly altered expressions. In place of (\ref{A24}) we have the simpler expansion
\be
\ln\Sigma^{(I)}(z)\simeq\sum_{k=1}^{\infty}\tilde{d}_kz^{-k}\label{B4}
\ee
where the first few coefficients are given by
\bea
\tilde{d}_1&=&\frac{1}{8}(1-4\nu^2)\;,\label{B5a}\\
\tilde{d}_2&=&\frac{1}{16}(1-4\nu^2)\;,\label{B5b}\\
\tilde{d}_3&=&\frac{1}{384}(25-104\nu^2+16\nu^4)\;,\label{B5c}\\
\tilde{d}_4&=&\frac{1}{128}(13-56\nu^2+16\nu^4)\;,\label{B5d}\\
\tilde{d}_5&=&\frac{1}{5120}(1073-4748\nu^2+1840\nu^4-64\nu^6)\;,\label{B5e}\\
\tilde{d}_6&=&\frac{1}{192}(103-465\nu^2+216\nu^4-16\nu^6)\;.\label{B5f}
\eea
The remaining analysis of Appendix A follows as before with these new coefficients in place of the $d_k$, and ({\ref{B3}) used for $g_\nu(z)$. (This changes the result we called $A_1(s)$ in (\ref{A35}).)


\begin{thebibliography}{99}
\bibitem{Kaluza}
T. Kaluza, Sitzungsber. Preuss. Akad. Wiss. Berlin, Math. Phys. Kl. (1921) 966.

\bibitem{Klein}
O.~Klein, Z. Phys. {\bf 37} (1926) 895; Nature (London) {\bf 118} 516.

\bibitem{DeWitt}
B.~S.~DeWitt, in {\bf Relativity, Groups, and Topology}, edited by C.~DeWitt and B.~S.~DeWitt (Gordon and Breach, New York, 1964).

\bibitem{Witten}
E.~Witten, Nucl. Phys. B {\bf 186} (1981) 412.

\bibitem{Nahm}
W.~Nahm, Nucl. Phys. B {\bf 135} (1978) 149.

\bibitem{DuffNilssonPope}
M.~J.~Duff, B.~Nilsson, and C.~N.~Pope, Phys. Rep. {\bf 130} (1986) 1 .

\bibitem{Arkani}
N.~Arkani-Hamed, S.~Dimopoulos, G.~Dvali, Phys. Lett. {\bf B429} (1998) 263; I.~Antoniadis, N.~Arkani-Hamed, S.~Dimopoulos, G.~Dvali, Phys. Lett. {\bf B436} (1998) 257.

\bibitem{RS}
L.~Randall and R.~Sundrum, Phys.~Rev.~Lett. {\bf 83} (1999) 3370; ibid.  4690.

\bibitem{GW1}
W.~D.~Goldberger and M.~B.~Wise, Phys. Rev. {\bf D60} (1999) 107505.
        
\bibitem{DHR}
H.~Davoudiasl, J.~L.~Hewett, and T.~G.~Rizzo, Phys. Lett. {\bf B473} (2000)  43.

\bibitem{Pomarol}
A.~Pomarol, Phys. Lett. {\bf B486} (2000) 153.

\bibitem{Grossman}
Y.~Grossman and M.~Neubert, Phys. Lett. {\bf B474} (2000) 361.

\bibitem{Kitano}
R.~Kitano, Phys. Lett. {\bf B481} (2000) 39.

\bibitem{Chang}
C.~V.~Chang and J.~N.~Ng, Phys. Lett. {\bf B488} (2000) 390.

\bibitem{Gherghetta}
T.~Gherghetta and A.~Pomarol, Nucl. Phys. B {\bf 586} (2000) 141.

\bibitem{Changetal}
S.~Chang, J.~Hisano, H.~Nakano, N.~Okada, and M.~Yamaguchi, Phys. Rev. D {\bf 63} (2000) 084025.

\bibitem{DHR2}
H.~Davoudiasl, J.~L.~Hewett, and T.~G.~Rizzo, Phys. Rev. D {\bf 63} (2001) 
075004.

\bibitem{Huber}
S.~J.~Huber and Q.~Shafi, Phys. Rev. D {\bf 63} (2001) 045010.
      
\bibitem{Flachi}
A.~Flachi and D.~J.~Toms, Phys. Lett. {\bf  B491} (2000) 157.

\bibitem{GW2}
W.~D.~Goldberger and M. B.~Wise, Phys. Rev. Lett. {\bf 83} (1999) 4922.

\bibitem{CW}
 P.~Candelas and S.~Weinberg, Nucl. Phys. {\bf B237} (1984) 397;

\bibitem{Garriga}
J. Garriga, O. Pujol\`{a}s, and T. Tanaka, hep-th/0004109.

\bibitem{DJTPLB} 
D.~J.~Toms, Phys. Lett. {\bf B484} (2000) 149.

\bibitem{GR}
W. D. Goldberger and I. Z. Rothstein, Phys. Lett. {\bf B491} (2000) 339.

\bibitem{DJTgravity}
D.~J.~Toms, Phys. Lett. {\bf B129} (1983) 31.

\bibitem{Isham}
C.~J.~Isham, Proc. Roy. Soc. (London) A {\bf 362} (1978) 383.

\bibitem{DJTChalkRiver}
D.~J.~Toms, in {\bf An Introduction to Kaluza-Klein Theories}, edited by H. C. Lee (Wold Scientific, Singapore, 1984).

\bibitem{DJTrenorm}
D.~J.~Toms, Phys. Rev. {\bf D26} (1982) 2713.

\bibitem{ColemanWeinberg}
S.~Coleman and E.~J.~Weinberg, Phys. Rev. {\bf D7} (1973) 1888.

\bibitem{Les}
E.~Elizalde, S.~Leseduarte, and A.~Romeo, J. Phys. {\bf A26} (1993) 2409; S.~Leseduarte and A.~Romeo, J. Phys. {\bf A27} (1994) 2483.

\bibitem{Bordag}
M.~Bordag, E.~Elizalde, and K.~Kirsten, J. Math. Phys. {\bf 37} (1996) 895; M.~Bordag, K.~Kirsten, and J.~S.~Dowker, Commun. Math. Phys. {\bf 182} (1996) 371.


\end{thebibliography}
\end{document}